\documentclass[aps,prl,twocolumn]{revtex4} 

\begin{document}

{\bf Comment on ``Domain Structure in a Superconducting Ferromagnet''}\\ 

According to Faur\`{e} and Buzdin \cite{FB} in a
superconducting ferromagnet a domain structure with
a period small compared with the London penetration depth $\lambda$ can arise. They
claim that this contradicts the conclusion of Ref. \onlinecite{S} that ferromagnetic domain
structure  in the Meissner state of a superconducting ferromagnet is absent at equilibrium.
Actually, there is no contradiction: The results of Ref. \onlinecite{S} have only been misunderstood.

First of all it is necessary to properly define what is a ferromagnetic domain structure. A
distinctive feature of a ferromagnetic state is a nonzero average spontaneous magnetization $\vec M$ in a
{\em macroscopic} volume. This takes place even in a ferromagnet with domains, since in ferromagnets
the domain size $l$ is macroscopic. It depends on the size and shape of the sample and on the orientation of $\vec
M$ with respect to the sample surface. For example, in a ferromagnetic slab of
thickness $L$, but infinite in other directions, there are no domains if $\vec M$ is parallel to the slab
surface. But if $\vec M$ is normal to the surface the stripe domains of the macroscopic size $l \propto
\sqrt{L}$ appear at equilibrium  \cite{LL}. 

On the other hand, from the very beginning of studying the coexistence of the ferromagnetism and
superconductivity it was known that competition between ferromagnetism and superconductivity may lead to
structures with periodic variation of the $\vec M$ direction in space. The period of these structures
is determined by the intrinsic parameters of the material, is normally smaller than $\lambda$, and does not
depend on the size and shape of the sample. Appearance of this structure means that
ferromagnetism has lost competition with superconductivity and the ``superconducting ferromagnet'' is not a
ferromagnet in a strict sense: this is an antiferromagnetic structure with a large but finite period.
Various types of such structures were known:
cryptoferromagnet alignment of Anderson and Suhl \cite{AS}, spiral structure of Blount and Varma \cite{BV},  or
domain structure of Krey \cite{Krey}. One can find these and other references in the review  \cite{BB} cited in
Ref. \onlinecite{S}. The second paragraph in Ref.
\onlinecite{S} clearly emphasized the difference between the ferromagnetic
macroscopic domains and these structures (let us call them intrinsic domain structures) and specifically
warned that the paper addressed the case when the material is stable with respect to formation of intrinsic
domains.  

Faur\`{e} and Buzdin \cite{FB} considered the intrinsic domain structure, which was analyzed by
Krey \cite{Krey} more than 30 years ago. They rederived the structure parameters obtained by him.
The domain size given by Faur\`{e} and Buzdin in Eq. (7), $l
\sim \tilde w^{1/3} \lambda^{2/3}$, coincides with that given by Krey in his Eq. (30) (apart from
notations). Here $\tilde w \sim (K/2\pi M^2) \delta$, $K$ is the energy of the easy-axis anisotropy,
and $\delta$ is the domain-wall thickness. The condition for formation of this
structure obtained by Krey also coincides with that of Faur\`{e} and Buzdin: $\lambda > \tilde w$.
Thus in the limit $L\to \infty$ they obtained the intrinsic domain structure in the state which
is globally antiferromagntic.  The structure can appear in any sample whatever its demagnetization factors are, in
particular, in the slab of thickness $L$ independently of whether $\vec M$ is normal or parallel to the slab
plane. Certainly the results of Ref. \onlinecite{S} cannot be relevant for this state as clearly warned there.
Faur\`{e} and Buzdin claimed that  their results for thin slabs (small $L$) disagree with Ref.  \onlinecite{S}, though Ref. \onlinecite{S}  did not consider finite-$L$ corrections at all addressing (like  Refs. \onlinecite{LL,Krey}) only the  macroscopic limit, when $L$ exceeds any intrinsic scales (including $\lambda$) or any combination of them. Only then  the difference between intrinsic  domains and {\em macroscopic} domains has a clear meaning.

Though time and again Faur\`{e} and Buzdin  stressed contradiction to Ref. \onlinecite{S}, in
reality they confirmed its conclusion: If the superconducting ferromagnet is stable with respect
to formation of intrinsic domains, macroscopic domains also do not appear. They claim that the area of stability, for
which the analysis of Ref.
\onlinecite{S} is relevant, corresponds to ``the nonrealistic limit of vanishing $\lambda$''. In reality Krey's stability condition $ \lambda <
\tilde w \sim (K/2\pi M^2)\delta$ (not $ \lambda \ll \tilde w $ !) is not so
severe and allows the values of $\lambda$ essentially larger than the domain-wall width $\delta$.
Indeed, the ratio $K/2\pi M^2$, which is called the quality factor of the magnetic material, can be
rather high. This is required for various applications of magnetic materials 
\cite{MS}. The quality factor is especially high for weak ferromagnetism, which is the most probable case for
the coexistence of ferromagnetism and superconductivity.

\vspace{0.18cm}

\noindent E.B. Sonin \\Racah Institute of Physics, Hebrew University of
Jerusalem, Jerusalem 91904, Israel \\



\begin{thebibliography}{99}

\bibitem{FB} M. Faur\`{e} and A.I. Buzdin,  Phys. Rev. Lett
{\bf 94}, 187202 (2005).
\bibitem{S} E.B. Sonin,  Phys. Rev. B
{\bf 66}, 100504(R) (2002).
\bibitem{LL} L.D. Landau and E.M. Lifshitz, {\sl Electrodynamics of Continuous 
Media} (Pergamon Press, Oxford, 1984).
\bibitem{AS} P.W. Anderson and H. Suhl, Phys. Rev. {\bf 116}, 898 (1959).
\bibitem{BV}  E.L. Blount and C.M. Varma, Phys. Rev. Lett
{\bf 42}, 1079 (1979).
\bibitem{Krey} U. Krey, Intern. J. Magnetism, {\bf 3}, 65 (1972).
\bibitem{BB} L.~N.~Bulaevskii, A.~I.~Buzdin, M.~L.~Kuli\'c, and S.~V.~
Panjukov, Adv.~Phys. {\bf 34}, 176 (1985).
\bibitem{MS} A.P. Malozemoff and J.C. Slonczewski, {\sl Magnetic 
Domain Walls in Bubble Materials} (Academic Press, N.Y., 1979).
\end{thebibliography}
\end{document}